\documentclass[12pt]{article}
\usepackage{graphicx}
\usepackage{amsmath}
\usepackage{amsfonts}
\usepackage{amssymb}
\newtheorem{theorem}{Theorem}[section]

\newtheorem{lemma}[theorem]{Lemma}

\newtheorem{remark}[theorem]{Remark}
\newenvironment{proof}[1][Proof]{\textsc{#1.} }{\ \rule{0.5em}{0.5em}}
\numberwithin{equation}{section}

\begin{document}

\title{Half polarized $U(1)$ symmetric vacuum spacetimes with AVTD behavior}
\author{Yvonne Choquet-Bruhat$^{\ast}$ and James Isenberg$^{\ast\ast}$.}
\maketitle

\textbf{Abstract. }In a previous work, we used a polarization condition to
show that there is a family of $U(1)$ symmetric solutions of the vacuum
Einstein equations on $\Sigma\times S^{1}\times R$ ($\Sigma$ any 2 dimensional
manifold) such that each exhibits AVTD\footnote{{\footnotesize Asymptotic
Velocity Term Dominated}} behavior in the neighborhood of its singularity.
Here we consider the general case of $S^{1}$ bundles over the base
$\Sigma\times R$ and determine a condition, called the half polarization
condition, necessary and sufficient in our context, for AVTD\ behavior near
the singularity.

\section{Introduction.}

A rigorous study of the singularities in cosmological solutions of the vacuum
Einstein equations has been hampered by the fact that the generic such
solution is expected to have a singularity of the oscillatory type predicted
by Belinsky, Lifshitz and Khalatnikov [BLK]. There is currently no
satisfactory mathematical method for treating such oscillating singularities,
at least when the spacetimes under study are spatially inhomogeneous. For this
reason much effort has gone into the study of families of solutions which have
milder cosmological singularities such as those of AVTD (asymptotically
velocity term dominated) type, for which rigorous, so called Fuchsian, methods
are available.

To suppress the oscillatory behavior expected for the generic solution, one can

\begin{itemize}
\item (i) introduce suitable matter sources such as scalar fields and study
the solutions of the associated non vacuum field equations [AR],

\item (ii) study higher dimensional models motivated by string or supergravity
theories wherein (for sufficiently high dimensions at least) the oscillations
are naturally suppressed [DHRW]; or

\item (iii) remain in 3+1 dimensions but impose a combination of symmetry and
polarization conditions in order to achieve the desired AVTD behavior.
\end{itemize}

For the case of $U(1)$ symmetric vacuum solutions on the trivial $S^{1}$
bundle $T^{2}\times R\times S^{1}\rightarrow T^{2}\times R$ (with $U(1)$
symmetry imposed on the circular fibers) Isenberg and Moncrief [IM] have
showed, using Fuchsian methods, that AVTD behavior is achieved provided the
solutions considered are at least half polarized in a certain well defined
sense. The half polarization condition includes, as a special case, the fully
polarized solutions wherein the 3 planes orthogonal to orbits of the $U(1)$
isometry action are integrable and the vacuum 3+1 field equations reduce to a
system of 2+1 dimensional Einstein equations coupled to a massless scalar
field on the quotient manifold $T^{2}\times R.$ The more general (half
polarized) solutions admit, in addition, half the extra (asymptotic) Cauchy
data expected for a fully general, non polarized solution of the same ($U(1)$
symmetric) type. On the basis of numerical studies due to Berger and Moncrief
the fully general, non polarized $U(1)$ symmetric vacuum solution on
$T^{2}\times R\times S^{1}\rightarrow T^{2}\times R$ is expected to have an
oscillatory singularity and hence not to be amenable to Fuchsian analysis [BM].

Choquet-Bruhat, Isenberg and Moncrief have extended the analysis given in [IM]
to cover the case of polarized $U(1)$ symmetric vacuum solutions on manifolds
of the more general type $\Sigma^{2}\times R\times S^{1}\rightarrow\Sigma
^{2}\times R,$ where $\Sigma^{2}$ is an arbitrary compact surface and the
bundle (in view of the assumed polarization condition) is necessarily trivial.
In the present paper the polarization restriction is eliminated in favor of an
appropriate half polarization condition and the limitation to trivial $S^{1}$
bundles over the base $\Sigma^{2}\times R$ is also removed. The present work
thus demontrates the existence of a large family of vacuum $U(1)$ symmetric
solutions of half polarized type defined on trivial and non trivial bundles
over $\Sigma^{2}\times R$ (with $\Sigma^{2}$ an arbitrary compact surface) and
having AVTD singularity behavior. The half polarization condition used in [IM]
involved requiring one of the asymptotic functions to vanish. The half
polarization condition which we find here necessary and sufficient for
possible AVTD behavior can be understood in terms of the behavior of the VTD
solutions to which our solutions converge as one approaches the singularity.
Specifically a VTD solution is half polarized if and only if the set of
geodesics in the Poincar\'{e} plane which represent it (at different spatial
points) all tend to the same point as $t$ appoaches the singularity.

\section{Einstein equations.}

A spacetime metric on a manifold $V_{4}\equiv M\times R,$ with $M$ an $S^{1}$
principal fiber bundle over a surface $\Sigma,$ reads, if it is invariant
under the $S^{1}$ action on $V_{4},$
\begin{equation}
^{(4)}g\equiv e^{-2\phi}\text{ }^{(3)}g+e^{2\phi}(d\theta+A)^{2}%
\end{equation}
with $\theta$ a parameter on the (spacelike) circular orbit, $\phi$ a scalar,
$A$ a locally defined 1 - form and $^{(3)}g$ a lorentzian metric, all on
$V_{3}:=\Sigma\times R.$

The vacuum 3+1 Einstein equations $Ricci(^{(4)}g)=0$ for such an $S^{1}$
symmetric metric on $V_{4}$ are known [M86], [CB-M96] to be
equivalent\footnote{{\footnotesize If we choose an arbitrary harmonic 1 - form
appearing in the solution to be zero.}} to the wave map equation from
$(V_{3},^{(3)}g)$ into the Poincar\'{e} plane $P=:(R^{2},G),$ $\Phi\equiv
(\phi,\omega):$ $V_{3}\rightarrow R^{2},$ where
\begin{equation}
G\equiv2(d\phi)^{2}+\frac{1}{2}e^{-4\gamma}(d\omega)^{2},
\end{equation}
coupled to the 2+1 Einstein equations for $^{(3)}g$ on $V_{3}$ with the wave
map as the source field. The scalar function $\omega$ on $V_{3}$ is linked to
the differential \ $F$ of $A$ by the relation
\begin{equation}
d\omega=e^{4\phi}\ast F,\text{ \ \ with \ \ }F=dA.
\end{equation}
Thus in local coordinates $x^{\alpha}$, $\alpha=0,1,2,$ on $V_{3},$ with
$\eta$ the volume form of $^{(3)}g=^{(3)}g_{\alpha\beta}dx^{\alpha}dx^{\beta
},$ one has
\begin{equation}
F_{\alpha\beta}\equiv\frac{1}{2}e^{-4\phi}\eta_{\alpha\beta\lambda}%
\partial^{\lambda}\omega.
\end{equation}

The wave map equations are, with $^{(3)}\nabla$ the covariant derivative in
the metric $^{(3)}g$
\begin{equation}
g^{\alpha\beta}(^{(3)}\nabla_{\alpha}\partial_{\beta}\phi+\frac{1}{2}%
e^{-4\phi}\partial_{\alpha}\omega\partial_{\beta}\omega)=0
\end{equation}%
\begin{equation}
g^{\alpha\beta}(^{(3)}\nabla_{\alpha}\partial_{\beta}\omega-4\partial_{\alpha
}\omega\partial_{\beta}\phi)=0.
\end{equation}

The $2+1$ Einstein equations are, with ''.'' indicating a scalar product in
the metric $G$%
\begin{equation}
^{(3)}R_{\alpha\beta}=\partial_{\alpha}\Phi.\partial_{\beta}\Phi.
\end{equation}

To solve these equations we choose for $^{(3)}g$ a zero shift, we denote the
lapse by $e^{\lambda}$ and we weigh by $e^{\lambda},$ without restricting the
generality, the $t$ dependent space metric $g=g_{ab}dx^{a}dx^{b},$ $a,b=1,2.$
That is, we set
\begin{equation}
^{(3)}g\equiv-N^{2}dt^{2}+g_{ab}dx^{a}dx^{b}\text{ \ \ with \ \ }N\equiv
e^{\lambda},\text{ \ \ }g_{ab}\equiv e^{\lambda}\sigma_{ab},\text{\ \ }%
\end{equation}
We denote by $\sigma^{ab}$ the contravariant form of $\sigma.$ The extrinsic
curvature of $\Sigma_{t}$ in $(V_{3},^{(3)}g)$ is
\begin{equation}
k_{ab}:=-\frac{1}{2N}\partial_{t}g_{ab}\equiv-\frac{1}{2}(\sigma_{ab}%
\partial_{t}\lambda+\partial_{t}\sigma_{ab}).
\end{equation}
The mean extrinsic curvature $\tau$ is therefore
\begin{equation}
\tau:=g^{ab}k_{ab}\equiv-e^{-\lambda}(\partial_{t}\lambda+\frac{1}{2}\psi),
\end{equation}
where we have defined
\begin{equation}
\psi:=\sigma^{ab}\partial_{t}\sigma_{ab}.
\end{equation}

The connection coefficients (Christoffel symbols) of $^{(3)}g$ are found to be
(note that $^{(3)}g^{00}=-e^{-2\lambda},$ $^{(3)}g^{ab}=g^{ab}=e^{-\lambda
}\sigma^{ab})$
\begin{equation}
^{(3)}\Gamma_{ab}^{c}=\Gamma_{ab}^{c}(g)=\Gamma_{ab}^{c}(\sigma)+\frac{1}%
{2}(\delta_{b}^{c}\partial_{a}\lambda+\delta_{a}^{c}\partial_{b}\lambda
-\sigma^{cd}\sigma_{ab}\partial_{d}\lambda)
\end{equation}%
\begin{equation}
^{(3)}\Gamma_{00}^{0}=\partial_{t}\lambda\text{, \ \ }^{(3)}\Gamma_{0a}%
^{0}=\partial_{a}\lambda,\text{ \ \ }^{(3)}\Gamma_{00}^{a}=\sigma
^{ab}e^{\lambda}\partial_{a}\lambda,
\end{equation}%
\begin{equation}
^{(3)}\Gamma_{ab}^{0}=-e^{-\lambda}k_{ab},\text{ \ }^{(3)}\Gamma_{a0}%
^{b}=-e^{\lambda}k_{a}^{b}.
\end{equation}
In particular it holds that
\begin{equation}
^{(3)}g^{\alpha\beta}{}^{(3)}\Gamma_{\alpha\beta}^{0}=\frac{1}{2}\psi
e^{-2\lambda}.
\end{equation}
We see that the metric $^{(3)}g$ is in harmonic time gauge if and only if
$\psi=0.$

The Einstein equations split into constraints and evolution equations. We
denote by $S_{\beta}^{\alpha}\equiv^{(3)}R_{\beta}^{\alpha}-\frac{1}{2}%
\delta_{\beta}^{\alpha}{}^{(3)}R$ the Einstein tensor of $^{(3)}g,$ by
$T_{\beta}^{\alpha}$ the stress energy tensor of $\Phi,$ and we set
$\Sigma_{\beta}^{\alpha}\equiv S_{\beta}^{\alpha}-T_{\beta}^{\alpha}.$ The
constraints are:
\begin{equation}
C_{0}\equiv\Sigma_{0}^{0}\equiv-\frac{1}{2}\{R(g)-k.k+\tau^{2}-e^{-2\lambda
}\partial_{t}\Phi.\partial_{t}\Phi-g^{ab}\partial_{a}\Phi.\partial_{b}\Phi\}=0
\end{equation}
and (indices raised with $g^{ab},$ $\nabla$ the covariant derivative in the
metric $g)$%
\begin{equation}
C_{a}\equiv e^{\lambda}\Sigma_{a}^{0}\equiv-\{\nabla_{b}k_{a}^{b}-\partial
_{a}\tau+e^{-\lambda}\partial_{a}\Phi.\partial_{t}\Phi\}=0.
\end{equation}
The evolution equations are, with $N=e^{\lambda},$
\begin{equation}
N(^{(3)}R_{a}^{b}-\rho_{a}^{b})\equiv-\partial_{t}k_{a}^{b}+N\tau k_{a}%
^{b}-\nabla^{b}\partial_{a}N+NR_{a}^{b}-N\partial_{a}\Phi.\partial^{b}\Phi=0.
\end{equation}

In order to obtain a first order system in the Fuchsian analysis that we will
make, we introduce auxiliary unknowns $\Phi_{t},\Phi_{a},\sigma_{c}^{ab} $
which are identified with the first partial derivatives of $\Phi$ and the
covariant derivative of $\sigma$ with respect to a given $t$ independent
metric $\tilde{\sigma}.$ These new unknowns satisfy the evolution equations
\begin{equation}
\partial_{t}\Phi=\Phi_{t},
\end{equation}%
\begin{equation}
\text{ \ \ }\partial_{t}\Phi_{a}=\partial_{a}\Phi_{t},
\end{equation}%
\begin{equation}
\partial_{t}\sigma_{c}^{ab}=\tilde{\nabla}_{c}\partial_{t}\sigma^{ab}\text{
\ \ }%
\end{equation}
where, by the definitions of $\sigma$ and $k,$%
\begin{equation}
\partial_{t}\sigma^{ab}=2e^{2\lambda}k^{ab}+\sigma^{ab}\partial_{t}\lambda.
\end{equation}
The function $\lambda$ is not left unknown, but rather is determined by a
gauge condition from its VTD value.

\section{VTD equations and solutions.}

The Velocity Terms Dominated equations are obtained by dropping the space
derivatives in the equations.

We denote by a tilde quantities which are independent of $t,$ and we denote
VTD solutions using a hat.

\subsection{Einstein evolution VTD solutions.}

In order to obtain a global (on $\Sigma$) formulation we choose a VTD metric
which remains in a fixed conformal class over $\Sigma$ as $t$ evolves; we set
\begin{equation}
\hat{\sigma}_{ab}=\tilde{\sigma}_{ab}\text{ \ and \ \ }\hat{g}_{ab}%
=e^{\hat{\lambda}}\tilde{\sigma}_{ab}.
\end{equation}
Then
\begin{equation}
\hat{\psi}=0,\text{ \ \ }\hat{g}^{ab}=e^{-\hat{\lambda}}\tilde{\sigma}%
^{ab},\text{ \ \ }\partial_{t}\hat{g}_{ab}=e^{\hat{\lambda}}\tilde{\sigma
}_{ab}\partial_{t}\hat{\lambda},
\end{equation}
and the definition of $k$ gives that
\begin{equation}
\hat{k}_{ab}=-\frac{1}{2}\tilde{\sigma}_{ab}\partial_{t}\hat{\lambda},\text{
\ }\hat{k}_{a}^{b}=-\frac{1}{2}e^{-\hat{\lambda}}\delta_{a}^{b}\partial
_{t}\hat{\lambda},\text{ \ \ }\hat{\tau}:=\hat{k}_{a}^{a}=-e^{-\hat{\lambda}%
}\partial_{t}\hat{\lambda}.
\end{equation}

Requiring that these VTD quantities satisfy the VTD evolution equations, we
obtain
\begin{equation}
\partial_{t}\hat{k}_{a}^{b}=\hat{N}\hat{\tau}\hat{k}_{a}^{b}.
\end{equation}
Therefore, by straightforward computation
\begin{equation}
\partial_{tt}^{2}\hat{\lambda}=0;\text{ \ \ \ hence \ \ \ }\hat{\lambda
}=\tilde{\lambda}-\tilde{v}t
\end{equation}
with $\tilde{\lambda}$ and $\tilde{v}$\ arbitrary functions on $\Sigma,$
independent of $t.$ Then we have
\begin{equation}
\hat{k}_{ab}=\frac{1}{2}\tilde{v}\tilde{\sigma}_{ab},\text{ \ }\hat{k}_{a}%
^{b}=\frac{1}{2}\tilde{v}e^{-\hat{\lambda}}\delta_{a}^{b},\text{ \ \ }%
\hat{\tau}=e^{-\hat{\lambda}}\tilde{v}.
\end{equation}

\subsection{Wave map VTD solutions.}

The results for a VTD wave map are very different from the results obtained
for a scalar function [CBIM]. If we drop space derivatives in the wave map
equations we obtain geodesic equations in the target manifold, with $t$ the
length parameter on these geodesics so long as the 2+1 metric is in harmonic
time gauge. If we make the change of coordinates $Y=e^{2\phi}$\ in the target
(which defines a diffeomorphism from $R^{2}$ onto the upper half plane $Y>0),$
the metric $G$ takes a standard form for the metric of a Poincar\'{e} half
plane; namely
\begin{equation}
G\equiv\frac{1}{2}\{\frac{d\omega^{2}+dY^{2}}{Y^{2}}\},\text{ \ \ }Y=e^{2\phi
}.
\end{equation}
The VTD, geodesic, equations written in this metric read, with a prime
denoting the derivative with respect to $t$%
\begin{equation}
\omega^{\prime\prime}-2Y^{-1}\omega^{\prime}Y^{\prime}=0,
\end{equation}%
\begin{equation}
Y^{\prime\prime}+Y^{-1}\omega^{\prime}X^{\prime}=0.
\end{equation}
The general solution of these geodesic equations is represented in these
coordinates, as is well known, by half circles\footnote{{\footnotesize We
discard here the special case which corresponds to the polarized case, treated
elsewhere, where these circles are centered at infinity. The geodesics are
then the half lines X}$\equiv\omega=${\footnotesize constant,.}} centered on
the line $Y=0;$ specifically, with $A$ and $B$ arbitrary constants (that is,
independent of $t)$, the solution takes the form
\begin{equation}
\hat{\omega}=B+A\cos\theta,\text{ \ }\hat{Y}=A\sin\theta,\text{ \ \ }%
0<\theta<\pi.
\end{equation}
These functions $\omega$ and $Y$ satisfy the differential equations 3.8, 3.9
if and only if it holds that:
\begin{equation}
\frac{\theta^{\prime\prime}}{\theta^{\prime}}=\frac{\cos\theta}{\sin\theta
}\theta^{\prime}.
\end{equation}
Integrating this equation we have that, with $\tilde{w}$ independent of $t$
\begin{equation}
\theta^{\prime}=-\tilde{w}\sin\theta.
\end{equation}
Another integration gives that, with $\tilde{\Theta}$ independent of $t$
\begin{equation}
\tan\frac{\theta}{2}=\tilde{\Theta}e^{-\tilde{w}t}.
\end{equation}
If we now make the substitution $A=e^{2\tilde{\phi}}$ and $B=\tilde{\omega},$
then 3.10 reads
\begin{equation}
\hat{\phi}=\tilde{\phi}+\frac{1}{2}log(sin\theta),\text{ \ \ \ }\hat{\omega
}=\tilde{\omega}+e^{2\tilde{\phi}}cos\theta\text{\ }.\text{ \ \ }%
\end{equation}

\begin{remark}
The set of above formulas is identical to the following one
\begin{equation}
\hat{\omega}\equiv\tilde{\omega}+e^{2\tilde{\phi}}\frac{1-\tilde{\Theta}%
^{2}e^{-2\tilde{w}t}}{1+\tilde{\Theta}^{2}e^{-2\tilde{w}t}},\text{
\ \ }e^{2\hat{\phi}}\equiv\hat{Y}=e^{2\tilde{\phi}}\frac{2\tilde{\Theta
}e^{-\tilde{w}t}}{1+\tilde{\Theta}^{2}e^{-2\tilde{w}t}}.
\end{equation}
$\hat{Y}$ tends to zero when $t$ tends to $\infty,$ but $\hat{\omega}$ tends
to $\tilde{\omega}+e^{2\tilde{\phi}}.$
\end{remark}

\subsection{Einstein constraint VTD solutions.}

We deduce from 3.14 and 3.12 that
\begin{equation}
\partial_{t}\hat{\phi}\equiv\frac{1}{2}\hat{Y}^{-1}\hat{Y}^{\prime}=\frac
{1}{2}\frac{cos\theta\theta^{\prime}}{sin\theta}=-\frac{1}{2}\tilde
{w}cos\theta
\end{equation}
and
\begin{equation}
e^{-2\hat{\phi}}\partial_{t}\hat{\omega}=-\theta^{\prime}=\tilde{w}sin\theta.
\end{equation}

The Einstein VTD constraints reduce to the following equation
\begin{equation}
2\hat{C}_{0}\equiv-\hat{k}.\hat{k}+\hat{\tau}^{2}-e^{-2\hat{\lambda}%
}\{2(\partial_{t}\hat{\phi})^{2}+\frac{1}{2}e^{-4\hat{\phi}}(\partial_{t}%
\hat{\omega})^{2}\}=0.
\end{equation}
We have, using 3.16 and 3.17,
\begin{equation}
2(\partial_{t}\hat{\phi})^{2}+\frac{1}{2}e^{-4\hat{\phi}}(\partial_{t}%
\hat{\omega})^{2}=\frac{1}{2}\tilde{w}^{2}.
\end{equation}
We deduce therefore from 3.6 that the VTD constraint 3.18 is satisfied if and
only if
\begin{equation}
\tilde{v}^{2}=\tilde{w}^{2}%
\end{equation}

\section{Fuchsian expansion.}

\subsection{2+1 metric expansions.}

For the unknowns $\sigma$ and $k$ we choose the following expansions, with the
various $\varepsilon^{\prime}s$ being positive numbers to be chosen later
\begin{equation}
\sigma^{ab}=\tilde{\sigma}^{ab}+e^{-\varepsilon_{\sigma}t}u_{\sigma}^{ab}%
\end{equation}%
\begin{equation}
k_{a}^{b}=e^{-\lambda}(\frac{1}{2}\tilde{v}\delta_{a}^{b}+e^{-\varepsilon
_{k}t}u_{k,a}^{b}).
\end{equation}
Then
\begin{equation}
\tau\equiv k_{a}^{a}=e^{-\lambda}(\tilde{v}+e^{-\varepsilon_{k}t}u_{k,a}^{a}).
\end{equation}
We take as a gauge condition
\begin{equation}
\lambda=\hat{\lambda};\text{ \ \ \ hence \ \ }\partial_{t}\lambda=-\tilde{v}.
\end{equation}
Comparing the expressions 4.4 and 2.10 for $\tau,$ we find that this condition
is equivalent to the gauge fixing requirement
\begin{equation}
e^{-\varepsilon_{k}t}u_{k,a}^{a}+\frac{1}{2}\psi=0.
\end{equation}

\subsection{Wave map expansion.}

We expand $\Phi$ near its VTD value; that is we set
\begin{equation}
\phi=\hat{\phi}+e^{-\varepsilon_{\phi}t}u_{\phi}\text{ \ \ \ with
\ \ }\ \ \hat{\phi}=\tilde{\phi}+\frac{1}{2}log(sin\theta),
\end{equation}
while for $\omega,$ for convenience of computation, we choose to set
\begin{equation}
\omega=\hat{\omega}+e^{2\phi}e^{-\varepsilon_{\omega}t}u_{\omega}\text{
\ \ with \ \ \ }\hat{\omega}=\tilde{\omega}+e^{2\tilde{\phi}}cos\theta,
\end{equation}

\subsection{Expansion for first derivatives.}

We expand the auxiliary unknowns near the values of the derivatives of the VTD
solution. That is we set (see 3.16, 3.17)
\begin{equation}
\phi_{t}=\partial_{t}\hat{\phi}+e^{-\varepsilon_{\phi_{t}}t}u_{\phi_{t}}%
\equiv-\frac{1}{2}\tilde{w}cos\theta+e^{-\varepsilon_{\phi_{t}}t}u_{\phi_{t}},
\end{equation}%
\begin{equation}
\omega_{t}=\partial_{t}\hat{\omega}+e^{2\phi}e^{-\varepsilon_{\omega_{t}}%
}u_{\omega_{t}}\equiv e^{2\tilde{\phi}}\tilde{w}sin^{2}\theta+e^{2\phi
}e^{-\varepsilon_{\omega_{t}}}u_{\omega_{t}}%
\end{equation}
The expansions of $\phi_{a}$ and $\omega_{a}$ are defined similarly by
setting
\begin{equation}
\phi_{a}=\partial_{a}\hat{\phi}+e^{-\varepsilon_{\phi^{\prime}}}u_{\phi_{a}%
},\text{ \ \ \ }\omega_{a}=\partial_{a}\hat{\omega}+e^{2\phi}e^{-\varepsilon
_{\omega^{\prime}}t}u_{\omega_{a}}.
\end{equation}
We next compute $\partial_{a}\hat{\phi}$ and $\partial_{a}\hat{\omega}.$ It
follows from 3.14 that
\begin{equation}
\partial_{a}\hat{\phi}=\partial_{a}\tilde{\phi}+\frac{cos\theta}{2sin\theta
}\partial_{a}\theta,\text{ \ \ \ \ }\partial_{a}\hat{\omega}=\partial
_{a}\tilde{\omega}+e^{2\tilde{\phi}}(2cos\theta\partial_{a}\tilde{\phi
}-sin\theta\partial_{a}\theta).
\end{equation}
We compute $\partial_{a}\theta$ using 3.13 and elementary properties of sine
and cosine. We find that
\begin{equation}
\partial_{a}\theta=\tilde{\Theta}^{-1}sin\theta\partial_{a}(\tilde{\Theta
}-t\tilde{w})
\end{equation}
Therefore it holds that
\begin{equation}
\partial_{a}\hat{\phi}=\partial_{a}\tilde{\phi}+\tilde{\Theta}^{-1}%
\frac{cos\theta}{2}\partial_{a}(\tilde{\Theta}-\tilde{w}t).
\end{equation}
Then, writing $\partial_{a}\hat{\omega}$ as sum of a term independent of $t$
plus terms tending to zero when $t$ tends to infinity, we have
\begin{equation}
\partial_{a}\hat{\omega}\equiv\partial_{a}(\tilde{\omega}+e^{2\tilde{\phi}%
})-e^{2\tilde{\phi}}[2(1-cos\theta)\partial_{a}\tilde{\phi}-\tilde{\Theta
}^{-1}sin^{2}\theta\partial_{a}(\tilde{\Theta}-\tilde{w}t)]
\end{equation}

For $\sigma_{c}^{ab},$ since $\tilde{\nabla}_{c}\tilde{\sigma}^{ab}=0,$ we
set
\begin{equation}
\sigma_{c}^{ab}\equiv e^{-\varepsilon_{\sigma^{\prime}}t}u_{\sigma^{\prime}%
,c}^{ab}.
\end{equation}

\section{Fuchsian system for the evolution equations.}

Given the Fuchsian expansions of the previous section, the Einstein - wave map
evolution system reads as a first order system for the set of unknowns
$U\equiv(u_{\sigma},$ $u_{k},$ $u_{\Phi},$ $u_{\Phi_{t}},$ $u_{\Phi^{\prime
},\text{ }}u_{\sigma^{\prime}}).$

The differential system for $U$ is Fuchsian in a neighbourhood of $t=+\infty$
if it takes the form
\begin{equation}
\partial_{t}U-LU=e^{-\mu t}F(t,x,U,\tilde{\partial}U)
\end{equation}
with $L$ a linear operator independent of $t$ with non negative eigenvalues,
$\mu$ a positive number and $F$ a set of tensor fields linear in
$\tilde{\partial}U$, continuous in $t,$ analytic in $x$ and $U$ and uniformly
Lipshitzian in all its arguments in a neighbourhood of $U=0$, for $t$ large enough.

\subsection{Einstein evolution equations.}

\subsubsection{Equation for $u_{\sigma}.$}

The Fuchsian expansion 4.2 for $k$ yields the following equation:
\begin{equation}
\partial_{t}g^{ab}\equiv2Nk^{ab}\equiv2e^{\lambda}g^{ac}k_{c}^{b}\equiv
e^{-\lambda}(\tilde{v}\sigma^{ab}+2e^{-\varepsilon_{k}t}\sigma^{ac}u_{k,c}%
^{b}).
\end{equation}
Using $g^{ab}\equiv e^{-\lambda}\sigma^{ab}$ and $\partial_{t}\lambda
=-\tilde{v},$ we have
\begin{equation}
\partial_{t}g^{ab}\equiv e^{-\lambda}(\tilde{v}\sigma^{ab}+\partial_{t}%
\sigma^{ab}).
\end{equation}
Combining these equations together with the Fuchsian expansion of $\sigma$
results in the equation:
\begin{equation}
\partial_{t}u_{\sigma}^{ab}-\varepsilon_{\sigma}u_{\sigma}^{ab}%
=2e^{(\varepsilon_{\sigma}-\varepsilon_{k})t}\sigma^{ac}u_{k,c}^{b},
\end{equation}
which is of Fuchsian type if $\varepsilon_{k}>\varepsilon_{\sigma}>0.$

\subsubsection{Equation for $u_{k}.$}

The Fuchsian expansion of $k$ together with $N=e^{\lambda}$ and $\partial
_{t}\lambda=-\tilde{v}$ imply by straightforward computation that
\begin{equation}
\partial_{t}k_{a}^{b}\equiv e^{-\lambda}\{\frac{1}{2}\tilde{v}^{2}\delta
_{a}^{b}+e^{-\varepsilon_{k}t}(\tilde{v}-\varepsilon_{k})u_{k,a}%
^{b}+e^{-\varepsilon_{k}t}\partial_{t}u_{k,a}^{b}\},
\end{equation}
and
\begin{equation}
N\tau k_{a}^{b}\equiv e^{-\lambda}\{\frac{1}{2}\tilde{v}^{2}\delta_{a}%
^{b}+\tilde{v}e^{-\varepsilon_{k}t}u_{k,a}^{b})+e^{-\varepsilon_{k}t}%
u_{k,c}^{c}(\frac{1}{2}\tilde{v}\delta_{a}^{b}+e^{-\varepsilon_{k}t}%
u_{k,a}^{b})\}.
\end{equation}
We see that $e^{-\lambda}\tilde{v}^{2}$ disappears from the difference
$\partial_{t}k_{a}^{b}-N\tau k_{a}^{b},$ which motivates the choice of the
Fuchsian expansion.

To write the evolution equation 2.18 for $k$ we now compute
\begin{equation}
\nabla^{b}\partial_{a}N\equiv e^{-\lambda}\sigma^{bc}\nabla_{c}\partial
_{a}e^{\lambda}\equiv\sigma^{bc}[\partial_{c}\lambda\partial_{a}%
\lambda+\partial_{a}\partial_{c}\lambda-\Gamma_{ac}^{d}(g)\partial_{d}%
\lambda].
\end{equation}
On the other hand, since $\Sigma$ is 2 dimensional and $g$ is conformal to
$\sigma$ with a factor $e^{\lambda},$ we have that
\begin{equation}
NR_{a}^{b}\equiv e^{\lambda}R_{a}^{b}\equiv\frac{1}{2}e^{\lambda}\delta
_{a}^{b}R(g)=\frac{1}{2}\delta_{a}^{b}\{R(\sigma)-\Delta_{\sigma}\lambda\}.
\end{equation}
From these results, if we define
\[
f_{a}^{b}(t,u,u_{x}):=-\nabla^{b}\partial_{a}N+NR_{a}^{b}-N\partial_{a}%
\Phi.\partial^{b}\Phi
\]
then we calculate
\begin{equation}
f_{a}^{b}\equiv\sigma^{bc}[\partial_{c}\lambda\partial_{a}\lambda+\partial
_{a}\partial_{c}\lambda-\Gamma_{ac}^{d}(g)\partial_{d}\lambda]+\frac{1}%
{2}\delta_{a}^{b}[R(\sigma)-\Delta_{\sigma}\lambda]-\sigma^{bc}\Phi_{a}%
.\Phi_{c}%
\end{equation}
We see that $f_{a}^{b}$ is at most a second order polynomial in $t,$ is
analytic in $x$ when $\tilde{v},\tilde{w},\tilde{\lambda},\tilde{\sigma}$ are
analytic; is linear in $\partial u;$ and is analytic, bounded and Lipshitzian
in $u$ for $u$ bounded and for large\footnote{{\footnotesize This restriction
on t comes from the covariant components of }$\sigma${\footnotesize \ \ which
remain bounded as long as }$\sigma^{ab}${\footnotesize \ remains positive
definite. }} $t$, except eventually for the last term which reads
\begin{equation}
\sigma^{bc}\Phi_{a}.\Phi_{c}=2\sigma^{bc}(\phi_{a}\phi_{c}+\frac{1}%
{2}e^{-4\phi}\omega_{a}\omega_{c}).
\end{equation}
The expansion 4.10 of $\phi_{a}$ shows that it does not cause problems for the
boundedness of $f_{a}^{b}$. However the expansion of $\omega_{a}$ gives
\[
e^{-2\phi}\omega_{a}=e^{-2\phi}\partial_{a}(\tilde{\omega}+e^{2\tilde{\phi}%
})-e^{-2\phi+2\tilde{\phi}}[2(1-cos\theta)\partial_{a}\tilde{\phi}%
+\tilde{\Theta}^{-1}sin^{2}\theta\partial_{a}(\tilde{\Theta}-\tilde{w}t)]
\]%
\begin{equation}
+e^{-\varepsilon_{\omega^{\prime}}t}u_{\omega_{a}}.
\end{equation}
It follows from 4.6 that
\[
e^{2(\tilde{\phi}-\phi)}=\frac{e^{-2\delta\phi}}{sin\theta},\text{ \ \ with
\ \ }\delta\phi\equiv e^{-\varepsilon_{\phi}t}u_{\phi}.
\]
Therefore, using ($1-cos\theta)/sin\theta=tan(\theta/2)$ we have:
\begin{equation}
e^{-2\phi}\omega_{a}=e^{2\tilde{\phi}}\frac{e^{-2\delta\phi}}{sin\theta
}\partial_{a}(\tilde{\omega}+e^{2\tilde{\phi}})-e^{-2\delta\phi}%
[2tan\frac{\theta}{2}\partial_{a}\tilde{\phi}+\tilde{\Theta}^{-1}%
sin\theta\partial_{a}(\tilde{\Theta}-\tilde{w}t)]+e^{-\varepsilon
_{\omega^{\prime}}t}u_{\omega_{a}}.
\end{equation}
We see that $e^{-2\phi}\omega_{a}$ will increase like ($sin\theta)^{-1}$ -
that is, like $e^{\tilde{w}t}$ - as $t$ tends to infinity, except if
\begin{equation}
\tilde{\omega}+e^{2\tilde{\phi}}=constant.
\end{equation}
Condition 5.13 is a generalization of the condition imposed on the fields in
[IM], with other notations, to obtain AVTD behaviour, in the case that
$\Sigma$ is a torus. Following the terminology of [IM] we call equation 5.13
the ''half polarization'' condition. Its geometric meaning is that \textbf{the
set of geodesics in the Poincar\'{e} plane representing the VTD solution all
tend to the same point of the axis }$Y=0$ \textbf{as }$t$ \textbf{tends to infinity.}

After inserting the Fuchsian expansions and multiplying by $e^{\lambda
+\varepsilon_{k}:t}$ we find that the equation 2.18 takes the form
\begin{equation}
\partial_{t}u_{k,a}^{b}-\varepsilon_{k}u_{k,a}^{b}-\frac{1}{2}v\delta_{a}%
^{b}u_{k,c}^{c}=e^{-\varepsilon_{k}t}u_{k,c}^{c}u_{k,a}^{b}+e^{\lambda
+\varepsilon_{k}t}f_{a}^{b}(t,u,u_{x}).
\end{equation}
Since $\lambda=\tilde{\lambda}-\tilde{v}t$ and $\tilde{v}=\tilde{w}$ this
system can take a Fuchsian form only if the functions $\tilde{\omega}$ and
$\tilde{\phi}$ satisfy the half polarization condition 5.13. To obtain the
system in obviously Fuchsian form in that case, we split 5.14 into its trace
and its traceless parts. For the trace part we have
\begin{equation}
\partial_{t}u_{k,a}^{a}-\varepsilon_{k}u_{k,a}^{a}-\tilde{v}u_{k,a}%
^{a}=e^{-\varepsilon_{k}t}u_{k,c}^{c}u_{k,a}^{a}+e^{\lambda+\varepsilon_{k}%
t}f_{a}^{a}(t,u,u_{x}).
\end{equation}
This equation takes Fuchsian form if and only if 5.13 is satisfied and
$\tilde{v}>\varepsilon_{k}.$ The same is verified for the traceless part
$^{T}u_{k,a}^{b},$ which satisfies an equation with left hand side
\begin{equation}
\partial_{t}\text{ }^{T}u_{k,a}^{b}-\varepsilon_{k}{}^{T}u_{k,a}^{b}.
\end{equation}

\subsubsection{Equation for $u_{\sigma^{\prime}}.$}

Using the expansion of $k$ and the relation $\partial_{t}\lambda=-\tilde{v},$
we find that
\begin{equation}
\partial_{t}\sigma^{ab}=2e^{2\lambda}k^{ab}+\sigma^{ab}\partial_{t}%
\lambda=2e^{-\varepsilon_{k}t}\sigma^{ac}u_{k,c}^{b}%
\end{equation}
The equation for $\sigma_{c}^{ab}$ gives therefore the following equation for
$u_{\sigma^{\prime}\text{ }}:$%
\begin{equation}
\partial_{t}u_{\sigma^{\prime},c}^{ab}-\varepsilon_{\sigma^{\prime}}%
u_{\sigma^{\prime},c}^{ab}=2e^{(\varepsilon_{\sigma^{\prime}}-\varepsilon
_{k})t}\tilde{\nabla}_{c}[\sigma^{ac}u_{k,c}^{b}.]\text{ }%
\end{equation}
which is of Fuchsian type so long as $\varepsilon_{\sigma^{\prime}%
}<\varepsilon_{k}.$

\subsection{Wave map equations.}

\subsubsection{Equations for auxiliary variables.}

The equations resulting from the introduction of the new variables $\phi
_{t},\omega_{t}$ are
\begin{equation}
\partial_{t}\phi-\phi_{t}=0,\text{ \ \ \ \ }\partial_{t}\omega-\omega_{t}=0.
\end{equation}
The first equation is of Fuchsian type for $u_{\phi}$ if $\varepsilon
_{\Phi_{t}}>\varepsilon_{\Phi\text{ }},$ since it reads
\begin{equation}
\partial_{t}u_{\phi}-\varepsilon_{\phi}u_{\phi}=e^{(-\varepsilon_{\Phi_{t}%
}+\varepsilon_{\Phi})t}u_{\phi_{t}}.
\end{equation}
The second equation reads
\begin{equation}
\lbrack\partial_{t}u_{\omega}+(2\phi_{t}-\varepsilon_{\omega})u_{\omega
}]\text{ }-e^{(\varepsilon_{\Phi}-\varepsilon_{\Phi_{t}})t}u_{\omega_{t}}=0.
\end{equation}
We replace $\phi_{t}$ by its value \ given in 4.8, which we write as follows
\begin{equation}
\phi_{t}=-\frac{1}{2}\tilde{w}+\frac{1}{2}\tilde{w}(1-cos\theta
)+e^{-\varepsilon_{\Phi_{t}}t}u_{\phi_{t}}.
\end{equation}
Since $1-cos\theta$ falls off to zero as $e^{-2\tilde{w}t},$ the equation 5.21
is of Fuchsian type for $u_{\omega}$ if $\tilde{w}>0$ and $\varepsilon
_{\Phi_{t}}>\varepsilon_{\Phi}.$

In the equations 2.20 to be satisfied by $\phi_{a}$ and $\omega_{a},$ the
derivatives of the VTD terms disappear, due to the commutation of partial
derivatives. The equation for $\phi_{a}$ reads
\begin{equation}
\partial_{t}u_{\phi_{a}}-\varepsilon_{\phi^{\prime}}u_{\phi_{a}}%
=e^{-(\varepsilon_{\Phi_{t}}-\varepsilon_{\Phi^{\prime}})t}(\partial
_{a}u_{\phi_{t}}-t\partial_{a}w),
\end{equation}
while the equation for $\omega_{a}$ becomes, using the expressions for
$\omega_{t}$ and $\omega_{a}$
\begin{equation}
\partial_{t}u_{\omega_{a}}+(2\phi_{t}-\varepsilon_{\omega^{\prime}}%
)u_{\omega_{a}}=e^{-(\varepsilon_{\Phi_{t}}-\varepsilon_{\Phi^{\prime}}%
)t}(\partial_{a}u_{\omega_{t}}+2\phi_{a}u_{\omega_{t}}).
\end{equation}
These equations are of Fuchsian type so long as $\tilde{w}>0$ and
$\varepsilon_{\Phi_{t}}>\varepsilon_{\Phi^{\prime}}.$

\subsubsection{Equation for $u_{\Phi_{t}}.$}

The first equation, 2.5, for the wave map reads
\[
g^{\alpha\beta}(\nabla_{\alpha}\partial_{\beta}\phi+\frac{1}{2}e^{-4\phi
}\partial_{\alpha}\omega\partial_{\beta}\omega)\equiv
\]%
\begin{equation}
-e^{-2\lambda}(\partial_{t}\phi_{t}+\frac{1}{2}e^{-4\phi}\omega_{t}\omega
_{t})+e^{-\lambda}\sigma^{ab}(\nabla_{a}\phi_{b}+\frac{1}{2}e^{-4\phi}%
\omega_{a}\omega_{b})+g^{\alpha\beta}\Gamma_{\alpha\beta}^{0}\phi_{t}=0
\end{equation}

Using the Fuchsian expansions for $\phi_{t}$ and $\omega_{t}$ together with
$\theta^{\prime}=-\tilde{w}\sin\theta$ and the value given in section 5.1.2
for $e^{2(\tilde{\phi}-\phi)}$we find that:
\[
\partial_{t}\phi_{t}+\frac{1}{2}e^{-4\phi}\omega_{t}\omega_{t}\equiv
\]%
\[
e^{-\varepsilon_{\Phi_{t}}t}(\partial_{t}u_{\phi_{t}}-\varepsilon_{\phi_{t}%
}u_{\phi_{t}})-\frac{1}{2}\tilde{w}^{2}\sin^{2}\theta+\frac{1}{2}%
(e^{-2\delta\phi}\tilde{w}sin\theta+e^{-\varepsilon_{\omega_{t}}t}%
u_{\omega_{t}})^{2}%
\]

On the other hand, using the expansions for $\sigma^{ab},\phi_{a}$ and
$\omega_{a}$ we find that:
\[
e^{-\lambda}\sigma^{ab}(\nabla_{a}\phi_{b}+\frac{1}{2}e^{-4\phi}\omega
_{a}\omega_{b})\equiv
\]%
\[
e^{-\lambda}(\tilde{\sigma}^{ab}+\delta\sigma^{ab})\{\nabla_{b}\partial
_{a}\hat{\phi}+e^{-\varepsilon_{\Phi^{\prime}}t}\nabla_{b}u_{\phi_{a}}%
+\frac{1}{2}(e^{-2\phi}\partial_{a}\hat{\omega}+e^{-\varepsilon_{\omega
^{\prime}}t}u_{\omega_{a}})(e^{-2\phi}\partial_{b}\hat{\omega}+e^{-\varepsilon
_{\omega^{\prime}}t}u_{\omega_{b}})\}
\]
We recall that
\[
\nabla_{b}\partial_{a}\hat{\phi}\equiv\nabla_{b}[\partial_{a}\tilde{\phi
}+\frac{cos\theta}{2}\tilde{\Theta}^{-1}\partial_{a}(\tilde{\Theta}-\tilde
{w}t)],
\]
while under the half polarization assumption
\begin{equation}
\tilde{\omega}+e^{2\tilde{\phi}}=constant,
\end{equation}
the product $e^{-2\phi}\partial_{a}\hat{\omega}$ is given by
\[
e^{-2\phi}\partial_{a}\hat{\omega}=-e^{-2\delta\phi}\tilde{\Theta}%
^{-1}sin\theta\partial_{a}(\tilde{\Theta}-\tilde{w}).
\]
Finally we calculate
\begin{equation}
g^{\alpha\beta}\Gamma_{\alpha\beta}^{0}\equiv\frac{1}{2}\psi e^{-2\lambda
}=-e^{-2\lambda-\varepsilon_{k}t}u_{k,a}^{a}.
\end{equation}

Inserting these computations into the first wave map equation produces an
equation of the form
\begin{equation}
\partial_{t}u_{\phi_{t}}-\varepsilon_{\Phi_{t}}u_{\phi_{t}}=e^{-\mu t}%
f_{\phi_{t}}(x,t,u,\partial u)
\end{equation}
which is of the Fuchsian type 5.1 (with $\mu>0)$ so long as $\tilde
{v}>\varepsilon_{\Phi_{t}},$ and $\varepsilon_{k}>\varepsilon_{\Phi_{t}}.$

Analogous computations show that the equation for $u_{\omega_{t}}$ is Fuchsian
presuming these same inequalities hold.

\subsection{Results for evolution.}

As a consequence of the calculations above we have proven the following theorem.

\begin{theorem}
There exist a collection of positive numbers \{$\varepsilon_{\sigma
},\varepsilon_{\sigma^{\prime}},\varepsilon_{k},\varepsilon_{\Phi}%
,\varepsilon_{\Phi_{t}},\varepsilon_{\Phi^{\prime}}\}$ such that, given
analytic asymptotic data on $\Sigma,$ $\tilde{A}=\{\tilde{v}=\tilde{w},$
$\tilde{\lambda},$ $\tilde{\sigma},\tilde{\Theta},\tilde{\phi},\tilde{\omega
}\},$ the Einstein - wave map evolution system written in first order form for
the unknown $U,$ which defines $g,$ $k,$ $\Phi$ and auxiliary variables by the
Fuchsian expansions of section 4, is a Fuchsian system for $U$ if and only if
$\tilde{\phi}$ and $\tilde{\omega}$ satisfy the half polarisation condition
5.13 and $\tilde{v}>0.$ It admits then one and only one analytic solution
tending to zero at infinity.
\end{theorem}

To show that this result implies that we have a family of solutions of the
Einstein - wave map evolution system which decays to solutions of the VTD
equations, we need to verify that for a large enough $t$ we have $\Phi
_{t}=\partial_{t}\Phi,$ $\Phi_{a}=\partial_{a}\Phi$ and the like. To show that
$\Phi_{a}=\partial_{a}\Phi$ we use the equations 2.20 together with
commutation of partial derivatives to show that:
\begin{equation}
\partial_{t}(\phi_{a}-\partial_{a}\phi)=\partial_{a}\phi_{t}-\partial
_{a}\partial_{t}\phi=0;
\end{equation}
hence $\phi_{a}-\partial_{a}\phi$ is independent of $t.$ As $t$ tends to
$\infty$ it tends to zero because
\begin{equation}
\phi_{a}-\partial_{a}\phi=e^{-\varepsilon_{\phi^{\prime}}t}u_{\phi_{a}%
}-e^{-\varepsilon_{\phi}t}(\partial_{a}u_{\phi}-\varepsilon_{\phi}u_{\phi}).
\end{equation}
It must therefore always be zero. Analogous arguments can be used to show that
$\omega_{a}=\partial_{a}\omega$ and $\sigma_{c}^{ab}=\tilde{\nabla}_{c}%
\sigma^{ab}.$

\section{Constraints.}

The solution of the evolution system satisfies the full Einstein equations so
long as it satisfies also the Einstein constraints, that is
\[
C_{0}:=\Sigma_{0}^{0}\equiv-\frac{1}{2}\{R(g)-k.k+\tau^{2}-e^{-2\lambda
}\partial_{t}\Phi.\partial_{t}\Phi\}=0
\]%
\[
C_{a}:=e^{\lambda}\Sigma_{a}^{0}\equiv-\{\nabla_{b}k_{a}^{b}-\partial_{a}%
\tau+e^{-\lambda}\partial_{t}\Phi.\partial_{a}\Phi\}=0.
\]

As usual we will rely on the Bianchi identities, here to construct a Fuchsian
system satisfied by the constraints. Together with the wave equation satisfied
by $\Phi,$ the Bianchi identities imply that
\begin{equation}
^{(3)}\nabla_{\alpha}\Sigma_{\beta}^{\alpha}=0.\text{ }%
\end{equation}
Modulo the evolution equations $^{(3)}R_{a}^{b}-\rho_{a}^{b}=0$ that we have
solved, with $\rho_{a}^{b}\equiv\Phi_{a}.\Phi^{b},$ it holds that
\begin{equation}
^{(3)}R-\rho=R_{0}^{0}-\rho_{0}^{0};
\end{equation}
hence
\begin{equation}
\Sigma_{0}^{0}\equiv R_{0}^{0}-\rho_{0}^{0}-\frac{1}{2}\delta_{0}^{0}%
(^{(3)}R-\rho)=\frac{1}{2}\delta_{0}^{0}(^{(3)}R-\rho)
\end{equation}
and
\begin{equation}
\Sigma_{a}^{b}=-\frac{1}{2}\delta_{a}^{b}(^{(3)}R-\rho)=-\delta_{a}^{b}%
\Sigma_{0}^{0}%
\end{equation}
We use these equations and the identities
\begin{equation}
\Sigma_{a}^{0}\equiv e^{-\lambda}C_{a},\text{ \ }\Sigma_{0}^{a}\equiv
-N^{2}\Sigma^{a0}\equiv-g^{ab}N^{2}\Sigma_{b}^{0}\equiv-e^{\lambda}g^{ab}C_{b}%
\end{equation}
together with the expressions for the Christoffel symbols of the metric
$^{(3)}g.$ We find that the equations 6.1 can be written in the form
\begin{equation}
\partial_{t}C_{0}-2e^{\lambda}\tau C_{0}=g^{ab}\nabla_{a}(e^{\lambda}%
C_{b})+g^{ab}e^{\lambda}\partial_{a}\lambda C_{a}%
\end{equation}
and (after some simplifications and multiplying by $e^{\lambda})$%
\begin{equation}
\partial_{t}C_{a}-e^{\lambda}\tau C_{a}=e^{\lambda}\nabla_{a}C_{0}%
+2e^{\lambda}\partial_{a}\lambda C_{0}.
\end{equation}
Equivalently, we have
\begin{equation}
\partial_{t}(e^{2\lambda}C_{0})-2\tilde{v}e^{2\lambda}C_{0}-2e^{\lambda}\tau
e^{2\lambda}C_{0}=e^{\lambda}\sigma^{ab}\nabla_{a}(e^{\lambda}C_{b}%
)+\sigma^{ab}e^{\lambda}\partial_{a}\lambda e^{\lambda}C_{a}%
\end{equation}
and
\begin{equation}
\partial_{t}(e^{\lambda}C_{a})-\tilde{v}e^{\lambda}C_{a}-e^{2\lambda}\tau
C_{a}=\nabla_{a}(e^{2\lambda}C_{0}).
\end{equation}
We see that $e^{2\lambda}C_{0}$ and $e^{\lambda}C_{a}$ satisfy a linear
homogeneous system, which admits zero as a solution. This solution is the
unique one tending to zero at infinity, so long as the system is Fuchsian.

\begin{lemma}
The system 6.8, 6.9 is Fuchsian, for a solution of the evolution system, if
the VTD solution satisfies $\hat{C}_{0}=0$ (i.e. $\tilde{v}^{2}=\tilde{w}^{2}).$
\end{lemma}

\begin{proof}
Since the coefficients of the equations 6.8, 6.9\ are constructed from
solutions of the evolution system we may use the expansions and estimates
derived in previous sections. In particular we calculate
\begin{equation}
e^{\lambda}\tau\equiv\tilde{v}+e^{-\varepsilon_{k}t}u_{k,a}^{a}.
\end{equation}
Equation 6.8 can therefore be written as the following equation of Fuchsian
type:
\begin{equation}
\partial_{t}(e^{2\lambda}C_{0})-4\tilde{v}e^{2\lambda}C_{0}e^{2\lambda}%
C_{0}=e^{-\varepsilon_{k}t}u_{k,a}^{a}e^{2\lambda}C_{0}+e^{\lambda}\sigma
^{ab}\nabla_{a}(e^{\lambda}C_{b})+\sigma^{ab}e^{\lambda}\partial_{a}\lambda
e^{\lambda}C_{a}.
\end{equation}

Equation 6.9 is not a priori in Fuchsian form for the pair ($e^{\lambda}%
C_{a},e^{2\lambda}C_{0})$ in spite of the identity 6.10. However if we use the
identity
\begin{equation}
e^{2\lambda}C_{0}\equiv-\frac{1}{2}\{e^{2\lambda}R(g)-e^{2\lambda
}k.k+e^{2\lambda}\tau^{2}-\partial_{t}\Phi.\partial_{t}\Phi\}
\end{equation}
and the property $\hat{C}_{0}=0$ together with the expression for $R(g)$ given
in 5.9 we can show that there exists a number $\mu>0$ and a bounded function
$F(x,t)$ such that we have
\begin{equation}
|e^{2\lambda}C_{0}|\leq e^{-\mu t}F(x,t)\text{ \ and \ }|\partial
_{a}(e^{2\lambda}C_{0})|\leq e^{-\mu t}F(x,t).
\end{equation}
It follows that 6.9 takes Fuchsian form.
\end{proof}

\begin{theorem}
A solution of the evolution system satisfies the full Einstein wave map
equations if and only if the half polarized asymptotic data satisfies the
condition $\tilde{w}=\tilde{v},$ and also
\begin{equation}
\tilde{\Theta}=1\text{ \ \ and \ }\tilde{v}e^{-\tilde{\lambda}+2\tilde{\phi}%
}=constant.
\end{equation}
\end{theorem}

\begin{proof}
To complete the proof that $C_{0}=C_{a}=0$ it suffices to show that
$e^{2\lambda}C_{0}$ and $e^{\lambda}C_{a}$ tend to zero at infinity. We have
already checked that this is true for $e^{2\lambda}C_{0},$ as long as
$\tilde{w}=\tilde{v};$ i.e. $\hat{C}_{0}=0.$

We now study the asymptotic behaviour of $e^{\lambda}C_{a}.$ If we denote by
$\delta u$ the difference between a field $u$ and its VTD value, we calculate
(recall that $\lambda=\hat{\lambda},\hat{\sigma}=\tilde{\sigma})$
\begin{equation}
e^{\lambda}(C_{a}-\hat{C}_{a})\equiv e^{\lambda}\{(\nabla_{b}-\tilde{\nabla
}_{b})k_{a}^{b}+\tilde{\nabla}_{b}\delta k_{a}^{b}-\partial_{a}\delta
\tau\}+\delta(\Phi_{t}.\Phi_{a})
\end{equation}
with
\begin{equation}
\delta(\Phi_{t}.\Phi_{a})\equiv2\phi_{t}\delta\phi_{a}+2\hat{\phi}_{a}%
\delta\phi_{t}+\frac{1}{2}e^{-2\phi}\omega_{t}\delta(e^{-2\phi}\omega
_{a})+e^{-2\hat{\phi}}\hat{\omega}_{a}\delta(e^{-2\phi}\omega_{t}).
\end{equation}
We see that, in the half polarized case, the Fuchsian expansions imply that
$e^{\lambda}(C_{a}-\hat{C}_{a})$ tends to zero as $t$ tends to infinity. Using
the expressions for \ $\ \hat{k}_{a}^{b}$ and $\hat{\lambda},$ we see that
$e^{\hat{\lambda}}\hat{C}_{a}$ reads:
\[
e^{\hat{\lambda}}\hat{C}_{a}\equiv\frac{1}{2}e^{\hat{\lambda}}\partial
_{a}(e^{-\hat{\lambda}}\tilde{v})-\hat{\Phi}_{t}.\hat{\Phi}_{a}\equiv\frac
{1}{2}(\partial_{a}\tilde{v}-\tilde{v}\partial_{a}\tilde{\lambda}+\tilde
{v}\partial_{a}\tilde{v}t)-\hat{\Phi}_{t}.\hat{\Phi}_{a}%
\]
Using the expressions of $\hat{\lambda},\hat{\Phi}_{t},\hat{\Phi}_{a}$ and the
half polarization condition, we find after some computation that
\begin{equation}
\hat{\Phi}_{t}.\hat{\Phi}_{a}=-\tilde{w}\{cos\theta\partial_{a}\tilde{\phi
}+\frac{1}{2}\tilde{\Theta}^{-1}\partial_{a}(\tilde{\Theta}-\tilde{w}t)\}.
\end{equation}
Thus we find that the terms containing $t$ disappear from $e^{\hat{\lambda}%
}\hat{C}_{a}$ if $\tilde{v}=\tilde{w}$ and $\tilde{\Theta}=1.$ It follows that
$e^{\hat{\lambda}}\hat{C}_{a}$ tends to zero as $t$ tends to infinity (recall
that $cos\theta$ tends then to $1)$ if and only if
\[
\frac{1}{2}[\partial_{a}\tilde{v}-\tilde{v}\partial_{a}\tilde{\lambda}%
]-\tilde{v}\partial_{a}\tilde{\phi}=0,
\]
a condition equivalent to the hypothesis 6.14 of the theorem.
\end{proof}

\begin{remark}
In the half polarized case the VTD solution only satisfies asymptotically the
VTD momentum constraint, and only after being multiplied $e^{\hat{\lambda}}.$
\end{remark}

\textbf{Aknowledgements.}

We are grateful to Vincent Moncrief for interesting discussions about this paper.

We thank the Kavli Institute for Theoretical Physics at Santa Barbara,
L'Institut des Hautes Etudes Scientifiques at Bures-sur-Yvette and the
department of mathematics of the University of Washington for providing very
pleasant and stimulating environments for our collaboration on this work. This
work was partially supported by the NSF, under grants PHY-0099373 and
PHY-0354659 at Oregon.

\textbf{References.}

[BM] B. K. Berger and V.\ Moncrief ''Numerical evidence for an oscillatory
singularity in generic $U(1)$ symmetric cosmologies on $T^{3}\times R"$ Phys.
Rev. D 58 064023-1-8 (1998).

[M86] V.\ Moncrief Reduction of Einstein equations for vacuum spacetimes with
U(1) spacelike isometry group, Annals of Physics 167 (1986), 118-142

[CB-M 96] Y.\ Choquet-Bruhat and V.\ Moncrief Existence theorem for solutions
of Einstein equations with 1 parameter spacelike isometry group, Proc.
Symposia in Pure Math, 59, 1996, H.\ Brezis and I.E.\ Segal ed. 67-80.

[CBIM] Y. Choquet-Bruhat, J. Isenberg and V.Moncrief ''Topologically general
$U(1)$ symmetric Einsteinian spacetimes with AVTD\ behaviour'' Il Nuovo
Cimento B, Vol. 119, issue no. 7-9, 2005.

[IM] J. Isenberg and V. Moncrief, ``Asymptotic behavior in polarized and
half-polarized $U(1)$ symmetric spacetimes'', Class. Qtm. Grav. 19, 5361-5386 (2002).

[KR] S. Kichenassamy and A.D. Rendall, ``Analytical description of
singularities in Gowdy spacetimes'', Class. Qtm.Grav.15, 1339-1355 (1998).

[AR] L. Andersson and A. Rendall, ``Quiescent cosmological singularities'',
Comm. Math. Phys. 218, 479-511 (2001).

[DHRW] T. Damour, M Henneaux, A. Rendall, and M. Weaver, ``Kasner-like
behavior for subcritical Einstein-matter systems'', Ann. H. Poin. 3, 1049-1111 (2002).

[IM92] J. Isenberg and V, Moncrief, ``Asymptotic behavior of the gravitational
field and the nature of singularities in Gowdy spacetimes'', Ann. Phys. 99,
84-122 (1992).

[CBM] Y. Choquet-Bruhat and V.Moncrief, ``Future complete $U(1)$ symmetric
einsteinian spacetimes'', Ann. Henri Poincare, 2, 1007-1064 (2001) See also
``Non linear stability of einsteinian spacetimes with $U(1)$ isometry group'', gr-qc/0302021.

[CB] Y. Choquet-Bruhat, ``Future complete $U(1)$ symmetric einsteinian
spacetimes, the unpolarized case'', in ''50 Years of the Cauchy Problem'',
eds. P. Chrusciel and H. Friedrich (2004).

[K] S. Kichenassamy, ''Nonlinear Wave Equations'' (Dekker, NY) (1996).

*Acad\'{e}mie des Sciences, 23 quai Conti 755270 Paris cedex 06, France

**Department of Mathematics and Institute of Theoretical Science, University
of Oregon, Eugene, OR 97403-5203, USA
\end{document}